# Exploration of Low Numeric Precision Deep Learning Inference Using Intel® FPGAs

Philip Colangelo[1,2], Nasibeh Nasiri[1], Asit Mishra[1], Eriko Nurvitadhi[1], Martin Margala[2], Kevin Nealis[1]

[1] *Intel Corporation*
[2] *University of Massachusetts Lowell*
philip.colangelo@intel.com

*Abstract*—Convolutional neural networks (CNN) have been shown to maintain reasonable classification accuracy when quantized to lower precisions, however, quantizing to sub 8-bit activations and weights can result in classification accuracy falling below an acceptable threshold. Techniques exist for closing the accuracy gap of limited numeric precision networks typically by means of increasing computation. This results in a trade-off between throughput and accuracy and can be tailored for different networks through various combinations of activation and weight data widths. Customizable hardware architectures like FPGAs provide the opportunity for data width specific computation through unique logic configurations leading to highly optimized processing that is unattainable by full precision networks. Specifically, ternary and binary weighted networks offer an efficient method of inference for 2-bit and 1-bit data respectively. Most hardware architectures can take advantage of the memory storage and bandwidth savings that come along with a smaller datapath, but very few architectures can take full advantage of limited numeric precision at the computation level. In this paper, we present a hardware design for FPGAs that takes advantage of the bandwidth, memory, power, and computation savings of limited numerical precision data. We provide insights into the trade-offs between throughput and accuracy for various networks and how they map to our framework. Further, we show how limited numeric precision computation can be efficiently mapped onto FPGAs for both ternary and binary cases. Starting with Arria 10, we show a 2-bit activation and ternary weighted AlexNet running in hardware that achieves 3,700 images per second on the ImageNet dataset with a top-1 accuracy of 0.49. Using a hardware modeler designed for our low numeric precision framework we project performance most notably for a 55.5 TOPS Stratix 10 device running a modified ResNet-34 with only 3.7% accuracy degradation compared with single precision.

## I. INTRODUCTION

FPGAs have been carving their niche in deep learning applications from full precision accelerators taking advantage of Intel's Arria 10 [1] hard floating-point units [2] [19] to core logic designs exploiting low precision networks [9] [5]. Different approaches for utilizing FPGAs in CNN acceleration applications have been studied and only recently have they become competitive with state of the art machine learning accelerators [6]. The continued development of FPGA technologies offering larger and faster fabrics like Stratix 10 with 10 TFLOPS peak performance further help FPGAs secure their position as an industry standard choice. With the rise of lower numeric precision networks using both activation and weight data widths as low as 1-bit, FPGAs will shine by demonstrating their capabilities in this domain. One of the current problems however is that few frameworks exist that allow FPGAs to be easily benchmarked with a variety of low precision configurations for several different networks. Further, there is limited amounts of data available for CNN on new FPGA devices such as Stratix 10 to help motivate future research. To address these problems, we provide the following contributions:

- Extension to Intel DLA hardware and software stack to provide a framework capable of accelerating very low numeric precision networks with 8-bit and sub 8-bit activations and weights including ternary and binary.
- Projections for a state of the art Stratix 10 low numeric precision accelerator showing an array of configurations for varying processing element data widths.
- Comparison of Stratix 10 to Titan X Pascal GPU for different processing elements using ResNet-34, ResNet-50, and AlexNet topologies.

The rest of the paper is organized as follows: Section 2 provides the necessary background information related to our work, section 3 details our accelerator framework, and section 4 contains the evaluation and data gathered from our experiments along with discussion of the results.

## II. BACKGROUND

### A. Low Numeric Precision Deep Learning Inference

CNNs are built from stacking layers of primitive building blocks that include convolution, sub sampling, and typically a non-linearity activation function. It is within the convolution function that lies the filters used to weight a neurons input helping direct the flow of data within a network ultimately leading to probability distribution on the networks output. One of the more common use cases for CNN is image classification. In this type of application, an image is used as input to the network and the output probability distribution is across a group of image classes dependent on the data set. Traditionally, convolution filters were represented as double or single precision floating point values however deep learning research has proven that CNN's are resilient and can be trained down to very low numeric precisions [12] [13].

Quantization of data in a feed forward neural network occurs in two places: activations and weights. Further sections of this paper will detail the implications of quantizing one or the other or both but for now will cover the broader areas of interest. Comparing to single precision floating point (32-bits), quantizing activation and weight data to lower bits (e.g. 8-bits and below) results in many savings depending on the hardware accelerator being used. Savings include: memory footprint for weights and intermediate activations, bandwidth to read and write from memory, and lower power given less routing and hardware resources. Further, when quantizing to low bit widths the computation cost in hardware resources can go down due to lower level logic optimizations capable on certain accelerators.

Some hardware architectures such as GPUs can do dense 32-bit down to 8-bit multiplies and additions very efficiently and given the fact that lower bit precisions result in accuracy degradation, little motivation has existed for sub 8-bit inference. Motivation does exist, however, to map binary and ternary weighted networks onto FPGA because of the advantages gained in changing the computation from multiply accumulates to lower level logic operations. Traditional multiply accumulates can be costly compared to low bit core logic implementations because most computation (~90% seen across various networks) occurs in the convolution primitives resulting in very large processing arrays that require a lot of hardware resources. By changing the computation from DSP to core logic there is an opportunity to do more work more efficiently. As an example, in binary weighted networks, filters are quantized to be -1 or 1 represented in hardware as either 0 or 1. Shown in Figure 1. , a traditional multiplication in a binary weighted network can be reduced to a sign flip and mux, both of which can be implemented efficiently in the core logic of an FPGA. Ternary weighted networks optimize similarly, and binary networks in which both activations and weights are 1-bit, optimize down to an XNOR gate and pop count [17] instead of a multiplier and adder. Processing elements will be discussed in more detail in following sections.

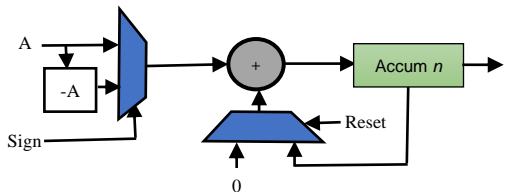

Figure 1.  Hardware implementations of low numeric precision networks can transform traditional multiply accumulate computation to simpler core logic operations.

There are several examples of research from academia and industry in reduced precision networks specifically focused on reducing the precision of weights for acceleration [5] [8] [14] [15]. Most of these networks sacrifice the accuracy over full precision networks yet techniques have begun to emerge that aim to reduce the accuracy gap created by quantizing data to lower widths. WRPN: Wide Reduced Precision Network [16] is an example of this type of research. The idea of WRPN being that accuracy lost by going to lower precision is recovered by increasing the number of filters in each layer. For instance, WRPN closes the accuracy gap of previously reported binary networks by increasing the depth of AlexNet 30%.

*B. Mapping to FPGA*

Modern FPGAs are made up of various building blocks useful for mapping an algorithm to its fabric. For example, Intel's Arria 10 mid-range device and Stratix 10 high performance device include hardened floating-point units in their DSP blocks capable of 1.5 TFLOPs and 10 TFLOPs peak performance respectively. Each DSP block can be configured for either floating-point or fixed-point arithmetic allowing for many applications. Various routing resources, registers, and core logic components come together to provide the framework our accelerator runs on.

Designing hand-tuned dot product engines is crucial for developing a high performant low bit accelerator. Traditional FP32 dense CNN FPGA accelerators use all DSP based computation however low bit networks can take advantage of the vast number of LUT based cells found in the FPGAs core logic to do multiply accumulates. Shown in Figure 2. is the high-level block diagram of Intel's Stratix 10 adaptive logic module (ALM), one of the basic building blocks used to create low bit dot product engines. ALMs provide up to 8 inputs that can be divided between two look up tables (LUT) and through various operating modes can provide a myriad of input and function combinations; please refer to [4] for more details. Further, DSPs and ALMs can be used in tandem to gain additional operations beyond using core logic alone, details will be provided in section 3.

Both Arria 10 and Stratix 10 devices contain on chip memory useful for caching filters and intermediate activations. RAM is presented as M20K blocks and can be configured for a given bit width. Further, several logic array blocks (LAB) containing ALM resources can be repurposed as memory logic array blocks (MLAB) providing dual-port SRAM for smaller memory needs such as creating FIFO's for pipelining. TABLE I. provides a breakdown of FPGA resources for two devices we considered during our experiments.

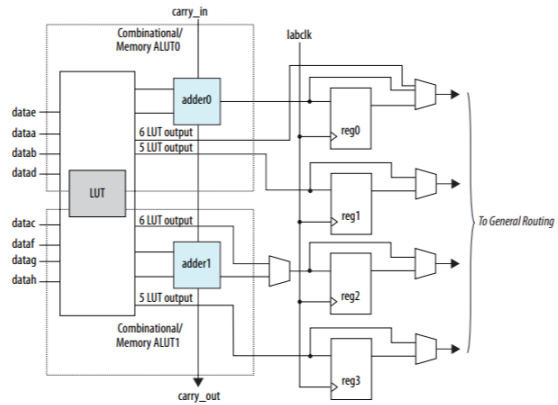

Figure 2.  Stratix 10 ALM Resource Diagram

TABLE I. FPGA RESOURCES

| | DSP | ALM | M20K | MLAB |
|---|---|---|---|---|
| Arria 10 GX 1150 | 1,518 | 427,200 | 54,260 (Kb) | 12,984 (Kb) |
| Stratix 10 GX 2800 | 5,760 | 933,120 | 229 (Mb) | 15 (Mb) |

*C. Related Work*

There have been many efforts in reducing the precision of weights and feature data since it can increase the classification throughput while providing savings in other areas such as memory, bandwidth, and power [26] [8] [29]. So far, the accuracy drop was inevitable by going down to lower precisions. For example, BinaryConnect [14] uses a single sign function to binarize the weights which are constrained to two values (e.g. -1 or 1). The authors of BinaryConnect benchmarked their reduced precision network on small data sets such as MNIST [18], CIFAR-10 [24] and SVHN [11]. Ternary weight networks [15] [25] is another work that attempts to reduce the precision of weights to save on total computation and shows less degradation on accuracy compared to binary weight networks. Finally, XNOR-Net [17] is another work which uses binarized weights and binarized activations to reduce computation in an extreme manner, but the top-1 accuracy drop for AlexNet is 12.5%. Few publications show support for sub 8-bit precision FPGA based accelerators [27] [9] [28], but these have been benchmarked with small datasets such as CIFAR-10 and SVHN which tend to retain accuracy for low precision when compared to larger data sets with more classes. Further, few papers have shown Stratix 10 performance for deep learning acceleration [6] and up till now, no publications existed that present 8-bit down to 1-bit activation and filter configurations that benchmark popular networks for a Stratix 10 based deep learning inference accelerator.

## III. THE FRAMEWORK

FPGAs have been making their mark in deep learning [6] [8][9] [10] by showing competitive inference implementations compared against GPUs. One of these FPGA implementations is Deep Learning Accelerator from Intel. DLA [19] is an OpenCL based framework that can exploit various levels of parallelism, cache intermediate data efficiently across a given network topology, optimize arithmetic operations, and maximize data reuse to manage DDR bandwidth. OpenCL allows the framework to be extensible in a more traditional software fashion compared to a hardware described design. For these reasons, we chose to use DLA as the base framework for our low bit implementations. We designed and implemented a convolutional neural network accelerator by modifying DLA to execute on a variety of activation and filter bit widths leading to a state of the art device. While the framework does allow for traditional bit widths such as 32, 16, and 8-bit, we focused on 8-bit and below to show the extraordinary capabilities of FPGA in this domain. In the following section, we will discuss the changes made to DLA in detail to support low bit inference.

*A. DLA Modifications and Optimizations*

DLA uses a fixed datapath architecture that contains various primitives and optimizations for traditional multiply accumulate convolutions. The first step in modifying this framework for low bit compute was to identify the blocks of the datapath that were not relevant to our experiments. Eliminating and modifying blocks from the datapath proved nontrivial because of the software centric flow of an OpenCL project. Figure 3. shows the high-level architecture for our accelerator. Each block is written in an OpenCL kernel and uses OpenCL channel extensions to communicate with other blocks.

One optimization for CNN introduced in [21] is the Winograd algorithm [20]. DLA utilizes this algorithm to reduce the total number of operations needed to produce an output in a convolution layer. Experiments showed that the Winograd algorithm breaks down for low bit compute because there is not enough information in the data to transform the inputs into the Winograd domain. Currently, there are no known algorithms to reduce the number of dot product operations for low bit width computations, so we remove the Winograd transformation blocks from the datapath.

Discussed in [15] [17] is the idea of training to reduce the Euclidean distance between full precision weights and either binary or ternary weights by including a scaling factor during training and inference. This optimization is used to derive low bit weight matrices used for inference and produce a positive per feature scaling factor to gain back accuracy lost from quantizing to such extremes. Essentially, for inference, this enables low bit optimized computation to occur for an entire feature map before scaling across with a full precision value. Because of the normalization scheme we chose (and will discuss next) we are able to hide this scalar in with other computation so we do not take additional operations to apply it.

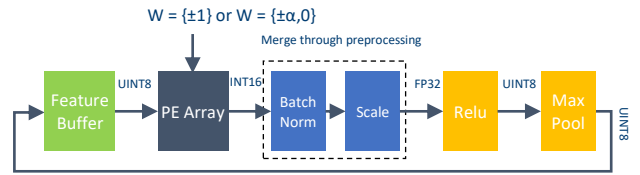

Figure 3. High level architecture for an 8-bit activation ternary or binary weighted accelerator based off DLA.

Normalization schemes such as local response normalization (LRN) are needed in CNN especially topologies using unbounded nonlinearities such as rectified linear unit (ReLU). Recall that ReLU passes through all positive activations and clips all negative activations to zero. Introducing normalization allows positive data to be bounded to not exceed the data container while preserving all necessary feature extraction information. LRN does follow a more neurobiological process as outlined in [30] by laterally inhibiting activations to bring out unique and large spiking neurons, however, low numeric precision networks favor more specialized normalization schemes such as Batch normalization (BN) [22]. BN has become a popular normalization method for reduced numeric precision networks because of its property of normalizing outputs across a mini batch to have a mean of zero and variance of one. While the original intent of [22] was to reduce the internal covariate shift allowing for faster training by reducing the effect each layer output has on the next, a positive side-effect of this process is that each layer is normalized as described above resulting in more effective quantization for low bit data. Along with BN is a scale kernel whose function is to train a set of parameters capable of representing the identity transform, i.e., train a new mean and variance. During training, BN derives its parameters from statistical information across mini batches while the scale kernel learns its parameters. During inference, these values can be merged together to form a single set of parameters. Favoring BN over LRN, we replace the traditional normalization scheme in the datapath with a fused Batch norm scale (BNS) primitive block that reads a single set of values per feature. Along with preprocessing the scale and shift parameters, we also merge these values with the alpha scale constant that is derived from training a ternary or binary network. Equations 1 and 2 show the process of merging the scales, shifts, and alpha scale together. Given the alpha scale as α, batch normalization scale and shift and scaling scale and shift as w, x, y, z respectively, and the merged scale and shift as γ and β respectively:

$$\gamma = \frac{y}{x} * \alpha \qquad (1)$$

$$\beta = z - \frac{y}{x} * w \qquad (2)$$

Quantizing output activations requires a method to represent accumulated data of a higher bit width down to a lower bit width without losing too much information. Data in our framework was quantized using the scheme found in [16] but with some minor optimizations. For the 2-bit activations and ternary weights (2xT) AlexNet proof of concept, the unoptimized quantizing scheme follows for an activation *x*:

$$q(x) = \frac{\text{floor}(\min(\max(0, x), 1)*3+0.5)}{3} \qquad (3)$$

Simply, from the output of batch normalization, FP32 activations are clipped between 0 and 1 and then stretched between 0 and 3. In practice, only values greater than 1 need to be clipped since we are quantizing the activations in ReLU which processes the *max*(0, x) function already. The data is then quantized to discrete values 0, 1, 2, and 3. In this state, hardware will represent these values as binary 00, 01, 10, and 11 but be interpreted as 0, 1/3, 2/3, and 1. Considering the optimized quantizing function:

$$q(x) = floor(\min(1, x) * 3 + 0.5) \qquad (4)$$

our quantization function becomes a clip and round with a multiplication.

DLA inherently uses the same data width and type for both weights and activations so we extended support to allow different widths and representations for activations and filters.

The datapath in our architecture varies as you propagate through the primitives. In any layer, the input activations to the PE are either an input image or the output from a previous layer which typically includes a non-linearity such as ReLU that clips all negative values to zero. In both cases, the input to the PE array will be an unsigned type. Accumulators inside the PEs must handle negative filter values so on the output of our array we use a signed data type that is wide enough to handle the input width accumulated by a dot product size. We carry this data through the merged batch norm and scale primitive where we apply single precision floating point scale and shift values to our activations. At this point, we clip all negative values to zero in ReLU and quantize our normalized data back to an unsigned data type.

Lastly, we created OpenCL Libraries which allow for hand tuned RTL modules to be pulled into the OpenCL design flow to implement the ternary and binary processing elements. In the next section, we will discuss these modules in detail.

*B. Processing Elements (PE)*

PEs are grouped together to form a 1D array of dot product engines. An array is capable of computing feature maps in parallel while each PE is designed to handle parallel computation across multiple dot product dimensions including input filter horizontal and vertical dimensions and output map horizontal and vertical dimensions. See [19] for an in-depth study of how DLA handles vectorization. While our low bit framework does not introduce additional dimensions of parallelism it is capable of handling greater magnitudes in each dimension because of its efficient PE core logic implementations.

Fully exploiting core logic capabilities for low bit width computations is demanding in a software environment because of the way algorithms are interpreted and mapped through the OpenCL compiler, so we used OpenCL libraries to develop fine-tuned dot product engines. OpenCL libraries, specific to Intel's OpenCL SDK for FPGAs [3], allow developers to write hardware description language (HDL) code that presents itself as an OpenCL function call allowing for a mix between software and hardware designs. Each PE configuration, i.e., the bit width of the inputs and activations and the bit width of the filters yields a different PE structure due to the architecture of the FPGA as described in the Mapping to FPGA section. Through individual cell instruction, ALMs are packed as efficiently as possible for each PE configuration leading to more available core logic to be used for greater levels of parallelism. Each PE is stamped out using the standard OpenCL flow from DLA to create a 1D systolic array of highly efficient hand-tuned PEs.

While efficient core logic PEs are optimal, there are available DSP resources that can provide additional performance to our accelerator. Using DSPs for their multipliers and a portion of the ALMs for accumulation, we can create more dot product engines for a total array that is greater than what core logic can provide alone. Each DSP block provides different operating modes that can be used for different data width configurations, so we will provide an example here based off one of our experiments and leave other configurations up to the reader. Given 2xT, an Arria 10 DSP block operating in fixed-point mode can be packed to provide 4x the compute density of a non-packed DSP. Using 18x18 mode, an Arria 10 DSP block provides two independent multipliers and outputs. Through packing, each multiplier can support four 2xT multiplies providing a total of eight 2xT multiplies per DSP block. Both independent outputs can then feed an ALM based adder tree as shown in Figure 4. Inputs 1a and 2a both contain packed data while 1b and 2b contain a single 2-bit data with padding on the sixteen MSBs. Packing four 2-bit data elements into 18-bits requires an additional sign extension and 2-bits padding in between data elements resulting in a total of 18-bits, see Figure 5. for a visual representation where $S_{ext}$ is sign extension, A, B, C, and D are data, and Pad is zero padding.

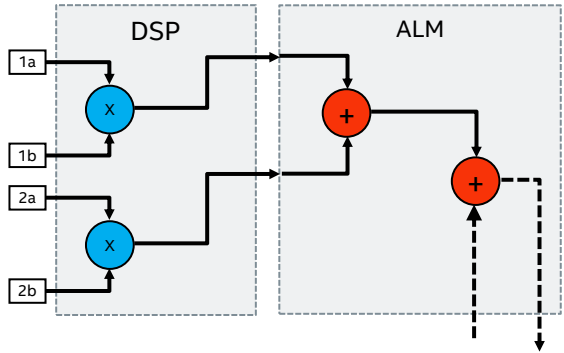

Figure 4. Arria 10 18x18 DSP mode allows for 4x packing of a 2-bit x 2-bit input.

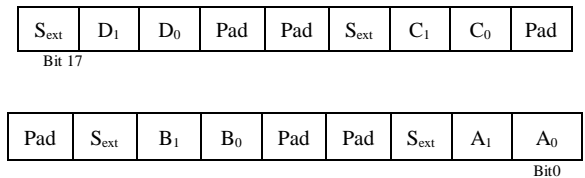

Figure 5. Four 2-bit data elements packed into an 18-bit DSP input

## IV. EVALUATION

We begin our evaluation and experiments by modelling our architecture to find the highest throughput at various PE configurations. Modelling allows us to search design spaces for highest throughput (TOP/sec) or efficiency, i.e., how well the PE array maps to a given topology. Various parameters and vectorizations go into design exploration and are detailed in [19]. Running HW configurations found by design exploration and verifying the validity of the model motivates the projections for other network configurations and devices. All accuracy results used in projections were taken from [16] and all experiments were run using the ImageNet dataset [7].

Throughout our experiments, we use a set of processing elements that range from 8-bits down to 1-bit for activation and weight input data. Each different processing element was RTL tuned for Stratix 10 logic. TABLE II. contains resource utilization data for each processing element configuration shown as ALMs/dot. For example, a single 8x8 processing element that has a dot product size of 8 (words/dot) uses 500 ALMs. Peak performance is directly related to the number of PEs a design can fit fueling reason to hand tune a smaller PE. The size of a PE in ALMs is related to the operation of that activation and weight bit width combination. Certain bit widths place and route differently than others due to the physical layout of an ALM and how many multiply accumulates a PE is designed to contain, i.e., certain bit widths resulted in a well packed PE giving high fit efficiency of the core logic.

TABLE II. PE CONFIGURATION LOGIC UTILIZATION

| Activation | Weight | words/dot | ALMs/dot |
|---|---|---|---|
| 8-bit | 8-bit | 8 | 500 |
| 8-bit | Ternary | 8 | 91 |
| 8-bit | Ternary | 16 | 176 |
| 8-bit | Binary | 8 | 77 |
| 8-bit | Binary | 16 | 149 |
| 8-bit | Binary | 32 | 298 |
| 4-bit | 4-bit | 8 | 210 |
| 4-bit | 4-bit | 16 | 431 |
| 3-bit | 3-bit | 8 | 70 |
| 2-bit | 2-bit | 8 | 39 |
| 2-bit | 2-bit | 16 | 91 |
| 2-bit | 2-bit | 64 | 437 |
| 2-bit | Ternary | 64 | 318 |
| 1-bit | 1-bit | 8 | 19 |
| 1-bit | 1-bit | 32 | 52 |

Experiments with an Arria 10 1150 were compiled and run in hardware to verify the modeler. Once there was a high confidence in searching the design space the modeler was modified further to support a Stratix 10 2800 device. Stratix 10 projections using the modeler were verified with partial design compiles including the PE array and compilation reports for individual kernels. Projections were made with an $f_{max}$ of 600 MHz.

## A. Accuracy and Throughput

Very low bit widths for activations and weights have been interesting for research but not deployment. One reason is because few architectures can take advantage of low bit computations and savings. Further, both GPUs and CPUs have traditionally been dominating the training and inference space, so little motivation has gone into exploring sub 8-bit deployment because of the lack of support on these architectures. Since FPGAs can take advantage of sub 8-bit optimizations, we consider these specialized PEs in conjunction with widening to gain back accuracy. Consider the curve in Figure 6. which shows accuracy vs throughput using three different widening schemes for the AlexNet topology. Data points with the highest throughput at a given accuracy were chosen between each widening scheme and PE configurations. Baseline AlexNet (1x) provides the highest throughput while doubling the number of filters (2x) provides higher accuracy. Choosing specific PE configurations with a widening scheme allows users to tailor the accuracy and throughput for their application.

Baseline AlexNet takes 1.44 GOPs to classify one image, so at FP32 activations and weights there are 64*1.44 or 92.16 GOP bits. Considering the 2xT example from above, there are 4*1.44 or 5.76 GOP bits resulting in a 16x savings in computation. Even at a 2x widening scheme which increases the number of GOPs by 4x (5.76 GOPs) resulting in 23.04 GOP bits, there is still a 4x savings in computation compared to the baseline FP32. Widening by 2x at 2xT results in a top-1 accuracy of approximately 56% which is only about 1% away from the FP32 baseline. This example provides insight into how low numeric precision PEs on an FPGA can be extremely efficient and when combined with techniques such as network widening, can provide better performance overall when compared to the baseline.

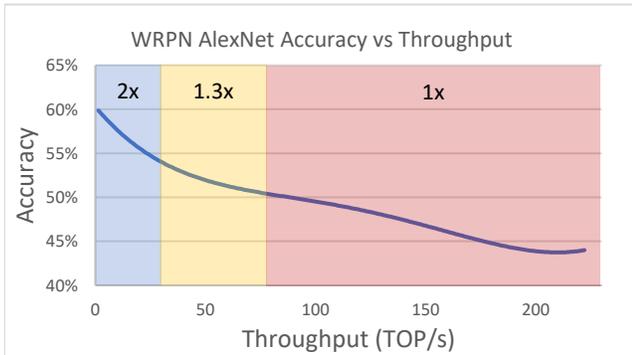

Figure 6.  Accuracy vs throughput curve for various widening schemes on AlexNet

## B. 2-bit Activations and Ternary Weighted AlexNet on Intel Arria 10 FPGA

We used the 2-bit activation and ternary weighted (2xT) AlexNet Arria 10 GX 1150 projections for our proof of concept device. 2xT 1x-wide AlexNet provides a top-1 accuracy of 49%. Running the performance modeler to search for an optimal solution for AlexNet, we arrived at an accelerator capable of 4.9 TOPs. Compiling and executing this design resulted in the following TABLE III. showing measured performance with accelerator information including $f_{max}$ and logic utilization. These results show that the performance modeler does a good job of determining a designs throughput based on the vectorizations and available logic for a specified PE configuration.

TABLE III.   ALEXNET 2xT HARDWARE RESULTS

| Device | Arria 10 GX 1150 |
|---|---|
| $f_{max}$ (MHz) | ~275 |
| ALM | 150,000 |
| M20K | 5000 |
| Throughput (Images/sec) | 3700 |

## C. Throughput vs Accuracy for ResNet on Intel Stratix 10 FPGA at Various Activation and Filter Data Widths

In this section, we evaluate different configurations of ResNet [23] as it maps to the PEs from TABLE II. ResNet has different flavors that vary in depth such as ResNet-34, 50, 101, and 152 each having more layers than the previous. The deeper the ResNet configuration, the more accurate the network is, i.e., ResNet-34 has an accuracy of ~74% while ResNet-152 has an accuracy of 78.6%. As we evaluate each PE we observe a basic trend of accuracy degradation as the PE input width is reduced, as expected. WRPN shows that networks which are widened, i.e., have deeper filters resulting in more output feature maps, are more accurate than the baseline due to the increase in number of activations ala deeper networks. Using this methodology, we evaluate ResNet-34 1x-wide (baseline), ResNet-34 2x-wide, ResNet-34 3x-wide, and ResNet-50 as a comparison to show tradeoffs between configurations. TABLE IV. shows a comprehensive evaluation of all ResNet flavors including throughput and accuracy for each PE. The two-important metrics in this table include equivalent tera ops per second (Eq TOPS) and top-1 accuracy (Top-1 Acc). Eq TOPS are the normalized total operations per second to the baseline topology, i.e., for the increase in computation in 2x and 3x wide topologies, we divide the total achievable performance by 4 and 9 respectively. Accuracies not reported in [16] are labeled as NR.

Starting with the baseline, single precision performance gives a top 1 accuracy of ~74% which then degrades as we lower the bit widths of our PE inputs. The trend in this table shows that there is a trade-off between accuracy and performance. While wider PEs give the best accuracy, they also have the lowest performance, and on the other end, shallower PEs provide the best performance but with more significant drops in accuracy. The reason for the performance increases are because lower bit widths use less logic freeing up resources for more processing elements. Further, very low bit width computation can be done more efficiently by reducing multiply accumulates to just mux and bit manipulations.

TABLE IV. also shows why low bit deep learning inference acceleration is a motivating factor for continued research. We noticed interesting trends in the data across the different ResNet widths pertaining to throughput vs accuracy. For the same (or close to) accuracy across widths there are different throughput projections. Looking at ResNet 34 baseline, 8x8 resulted in 6.55 TOPS device at 70.9% accuracy whereas ResNet34 3x wide 1x1 resulted in 24.7 TOPS at 72.4% accuracy. Of course, in the 3x wide version more calculations per image are needed but because we can run those calculations so efficiently in FPGA logic the result is a higher throughput compared to a less efficient, higher bit width version. It is the higher number of operations that close the accuracy gap and the lower bit widths that provide more throughput. The amount of data presented in this section provides a comprehensive study on how one can choose a configuration for a given network based on accuracy requirements. If throughput is met with less logic, then there are power savings or available logic for other work.

### D. Stratix 10 Inference Accelerator Comparison against Titan X Pascal

Using the low numeric precision modeler to search the configuration space, we find optimal solutions for our accelerator and use this data to compare against a state of the art device. TABLE V. provides the comparison between the chosen Stratix 10 2800 (S10) based accelerators and a Titan X Pascal GPU (TX). We evaluate performance in images per second for ResNet-34, ResNet-50, and AlexNet. Current GPU architectures can at best take advantage of 8-bit computation, any data widths less than 8-bit will be padded and run on an 8-bit architecture. GPUs typically perform best at higher batch sizes so we provide data for both batch size 1 (b1) and 128 (b128). Stratix 10 performance is affected very minimally by batch size showing that our architecture is outstanding in low (and deterministic) latency and low batch applications. While low batch sizes are optimal for this architecture, we can further optimize the accelerator for larger batch sizes.

Single PE array's use parallelism across a single image with average efficiency mapping across networks typically in the range of 50%-70%. Instead of searching for best performance, the modeler can search for best efficiency leading to smaller arrays mapping better than 90%. Stamping out these smaller arrays can provide an architecture capable of running multiple images in parallel. This becomes a trade-off between throughput and latency. Future work will include exploring this level of parallelism on the Stratix 10 device by further enhancing the framework to support this feature.

At high batch sizes, GPU has the highest overall throughput for larger data widths and with research showing 8-bit computations having very little accuracy degradation compared to FP32, GPU architectures now supporting 8-bit computations provide a nice linear throughput scaling making the bit width landscape more competitive. Still, our reduced precision FPGA accelerator can achieve the highest

TABLE IV. THROUGHPUT AND ACCURACY FOR VARIOUS PE CONFIGURATIONS ON RESNET TOPOLOGIES

| | | ResNet-34 1x Wide | | ResNet-34 2x Wide | | ResNet-34 3x Wide | | ResNet-50 | |
|---|---|---|---|---|---|---|---|---|---|
| *Activation* | *Weight* | *Eq TOPS* | *Top-1 Acc* | *Eq TOPS* | *Top-1 Acc* | *Eq TOPS* | *Top-1 Acc* | *Eq TOPS* | *Top-1 Acc* |
| FP32 | FP32 | 7 | 0.7359 | NR | NR | NR | NR | 7 | 0.7622 |
| 8-bit | 8-bit | 8 | 0.7093 | 2 | NR | 1 | NR | 8 | 0.7243 |
| 8-bit | Ternary | 43 | 0.6919 | 11 | NR | 5 | NR | 43 | 0.7038 |
| 8-bit | Binary | 52 | NR | 13 | NR | 6 | NR | 52 | NR |
| 4-bit | 4-bit | 18 | 0.7033 | 5 | 0.7453 | 2 | NR | 18 | 0.7188 |
| 3-bit | 3-bit | 51 | NR | 13 | NR | 6 | NR | 51 | NR |
| 2-bit | 2-bit | 85 | 0.6793 | 21 | 0.7332 | 9 | NR | 85 | NR |
| 2-bit | Ternary | 98 | 0.6793 | 25 | 0.7332 | 11 | NR | 98 | NR |
| 1-bit | 1-bit | 267 | 0.6054 | 67 | 0.6985 | 30 | 0.7238 | 267 | 0.6263 |

TABLE V. STRATIX 10 AND TITAN X COMPARISON IN IMAGES PER SECOND

| | | ResNet 34* | | | ResNet 50 | | | AlexNet | | |
|---|---|---|---|---|---|---|---|---|---|---|
| *Act.* | *Weight* | *S10 b1* | *TX b1* | *TX b128* | *S10 b1* | *TX b1* | *TX b128* | *S10 b1* | *TX b1* | *TX b128* |
| FP32 | FP32 | 470 | 435 | 1,214 | 448 | 415 | 1,156 | 2,400 | 823 | 5,882 |
| 8-bit | 8-bit | 535 | 590 | 3,977 | 509 | 562 | 3,787 | 2,730 | 972 | 18,714 |
| 8-bit | Ternary | 2,956 | 590 | 3,977 | 2,814 | 562 | 3,787 | 15,087 | 972 | 18,714 |
| 8-bit | Binary | 3,555 | 590 | 3,977 | 3,385 | 562 | 3,787 | 18,147 | 972 | 18,714 |
| 4-bit | 4-bit | 1,247 | 590 | 3,977 | 1,188 | 562 | 3,787 | 6,367 | 972 | 18,714 |
| 3-bit | 3-bit | 1,238 | 590 | 3,977 | 1,179 | 562 | 3,787 | 6,320 | 972 | 18,714 |
| 2-bit | 2-bit | 5,787 | 590 | 3,977 | 5,509 | 562 | 3,787 | 29,537 | 972 | 18,714 |
| 2-bit | Ternary | 4,885 | 590 | 3,977 | 4,651 | 562 | 3,787 | 24,933 | 972 | 18,714 |
| 1-bit | 1-bit | 10,073 | 590 | 3,977 | 9,591 | 562 | 3,787 | 51,417 | 972 | 18,714 |

*ResNet 34 Titan X numbers were acquired by scaling ResNet 50 down

throughput or lowest latency across networks given a complete design space. Of course, accuracy data must be considered when choosing a configuration for deployment and tables like TABLE IV. can provide a way of choosing the right one.

## V. CONCLUSIONS

In this paper, we presented data for a Stratix 10 FPGA based framework that can accelerate low numeric precision inference for deep learning. We showed that reduced precision networks typically incur a classification accuracy drop but can recover the gap compared to full precision through increased computation. This provides a multidimensional solution space between accuracy, throughput, and network computation size. Because of the reconfigurable logic of an FPGA, low precision compute can be exploited by performing bit precision operations. The coupling between low precision activations and filters and increased operations can provide unique solutions for accuracy and throughput tradeoffs compared to traditional approaches of benchmarking reduced precision networks. With this data, we compared our low precision accelerators images per second against a state of the art GPU. Our results show that FPGA based low numeric precision accelerators are state of the art devices that can perform in both low latency and high throughput deep learning applications.